\documentstyle[12pt,a4]{article}

\begin{document}

\baselineskip=18pt plus 0.2pt minus 0.1pt

\makeatletter

\def\p{{\partial}}
\def\nn{{\nonumber}}
\newcommand{\be}{\begin{equation}}
\newcommand{\ee}{\end{equation}}
\newcommand{\bea}{\begin{eqnarray}}
\newcommand{\eea}{\end{eqnarray}}
\newcommand{\tr}{\mathop{\rm tr}}
\newcommand{\Tr}{\mathop{\rm Tr}}
\newcommand{\Pf}{\mathop{\rm Pf}}
\renewcommand{\thefootnote}{\fnsymbol{footnote}}

\begin{titlepage}
\title{
\hfill\parbox{4cm}
{\normalsize KUNS-1728\\{\tt hep-th/0107203}}\\
\vspace{1cm}
USp(32) String\\as Spontaneously Supersymmetry Broken Theory
}
\author{
Sanefumi {\sc Moriyama}
\thanks{{\tt moriyama@gauge.scphys.kyoto-u.ac.jp}}
\\[7pt]
{\it Department of Physics, Kyoto University, Kyoto 606-8502, Japan}
}
\date{\normalsize July, 2001}
\maketitle
\thispagestyle{empty}

\begin{abstract}
\normalsize
It was suggested by Sugimoto that there is a new supersymmetry
breaking mechanism by an orientifold plane which is oppositely charged
as the usual one.
Here we prove the trace formula for this system to show that the
supersymmetry is broken not explicitly but spontaneously.
We also discuss the possibility of interpreting the orientifold plane
as an intrinsic object of the superstring theory.
\end{abstract}

\end{titlepage}

\section{Introduction and summary}
Since our world has no explicit supersymmetry, if we try to identify
the superstring theory as the unique theory of our world, the
supersymmetry in the superstring theory should be broken
spontaneously.
The supersymmetry breaking mechanism using both D-branes and
anti-D-branes or non-BPS D-branes has been fully investigated
\cite{BanSus,Sen,Yon}.
However, we still have little knowledge for a system whose
supersymmetry is broken by the orientifold plane.

Recently, it was suggested by Sugimoto \cite{Sug} that there is a new
supersymmetry breaking mechanism.
In his seminal paper he showed that it is possible to consider an
orientifold plane which is oppositely charged as the usual one.
In this case we should add 32 anti-D9-branes, instead of 32 D9-branes,
for cancellation of the Ramond-Ramond (R-R) D9-brane charge.
The resulting gauge group is USp(32) instead of SO(32) and the
supersymmetry in this system is completely broken.
The anomaly cancellation mechanism of this system is studied in
\cite{Sug,SW}.

In the present paper we shall prove explicitly the trace formula
\cite{FGP,WB,Yon}
\bea
\Tr_{\rm NS-R}m^2=0,
\eea
for this system.
Here the masses squared $m^2$ are summed over all the physical Hilbert
space with those of particles in the NS-sector (spacetime bosons)
contributing as they are and those of the R-sector particles
(spacetime fermions) with an extra negative sign.
In supersymmetric field theoretical models the degree of freedom of
bosons and that of fermions have to be balanced at every mass level.
If the supersymmetry is broken spontaneously, the degrees of freedom
are no longer balanced at every mass level.
Instead, the degrees of freedom are balanced as a whole in the sence
of the trace formula.
Hence, our result that the trace formula holds for this USp(32) string 
theory suggests that the supersymmetry is broken spontaneously.
The trace formula for the system with both D-branes and anti-D-branes
is first proved in \cite{Yon}.
Here we shall follow all the technics in it.
See also \cite{DMM} for general structures of the trace
formula.\footnote{The author is grateful to his colleagues for calling
his attention to the work. The work contains the trace formula for a
wide class of consistent string theories. Here, however, we shall
discuss more specific properties of the orientifold plane.}

This result is consistent with \cite{DudMou}.
In the paper the low energy effective supergravity theory for this
system is constructed as the type I theory \cite{CM}.
Although the closed string sector has explicit supersymmetry, the
supersymmetry of the open string sector is broken spontaneously and
realized non-linearly.

The motivation of the present paper is as follows.
Since we all believe that all the five perturbative superstring
theories are realized as particular configurations of the unique
M-theory, conceptually the orientifold plane in the type I theory
should also be realized as an intrinsic object in the off-shell
formalism of the superstring theory.
Although there are of course no off-shell formalisms to construct the
orientifold configuration explicitly, we would like to give some
evidences for this suggestion by proving the trace formula for a
non-supersymmetric system with the orientifold.
We shall return to this question in the final section.

The contents of the present paper are as follows.
In the next section we shall first review the spectrum of the USp(32)
string theory and then proceed to prove the trace formula for the
system.
The final section is devoted to conclusions and discussions.

\section{Trace formula in the USp(32) string theory}
In this section, we shall examine the trace formula for the USp(32)
string theory.
For this purpose, we shall first briefly review the open string
spectrum of the USp(32) theory \cite{Sug}.
Since the USp(32) theory is defined by reversing the orientifold
charge of the type I theory, let us begin by recalling the spectrum of
the type I theory.
The type I theory is obtained by projecting out worldsheet
orientation.
The R-R charge is cancelled only when we add 32 D9-branes.
The open string spectrum of the type I theory is summarized in the
partition function
\bea
Z=\Tr_{\rm NS-R}q^{-H}\Bigl(1+\Omega\Bigr)\Bigl(1+(-)^F\Bigr),
\eea
with $H=\alpha'm^2$ defined as
\bea
&&H_{\rm NS}=\sum_{m=1}^{\infty}\alpha_{-m}\cdot\alpha_m
+\sum_{r=1/2}^{\infty}rb_{-r}\cdot b_r-\frac12,\\
&&H_{\rm R}=\sum_{m=1}^{\infty}\alpha_{-m}\cdot\alpha_m
+\sum_{m=1}^{\infty}md_{m}\cdot d_m,
\eea
for the NS-sector and the R-sector, respectively.
Hereafter we shall split the partition function into the bosonic part
$Z_{\rm NS}$ and the fermionic part $Z_{\rm R}$:
$Z=Z_{\rm NS}-Z_{\rm R}$.
Each part is given as follows \cite{string}.
\bea
&&Z_{\rm NS}=(32)^2\cdot\frac{\vartheta^0_0(it)^8}{\eta(it)^8}
-(32)^2\cdot\frac{\vartheta^0_1(it)^8}{\eta(it)^8}
-32\cdot\frac{\vartheta^0_1(2it)^8\vartheta^1_0(2it)^8}
{\eta(2it)^8\vartheta^0_0(2it)^8},\label{ZNS}\\
&&Z_{\rm R}=(32)^2\cdot\frac{\vartheta^1_0(it)^8}{\eta(it)^8}
-32\cdot\frac{\vartheta^0_1(2it)^8\vartheta^1_0(2it)^8}
{\eta(2it)^8\vartheta^0_0(2it)^8}\label{ZR},
\eea
where the eta function $\eta(it)$ and the theta functions
$\vartheta^\alpha_\beta(it)$ are defined as
\bea
&&\eta(it)=q^{1/24}\prod_{m=1}^{\infty}(1-q^m),\\
&&\vartheta^0_0(it)=q^{-1/48}\prod_{m=1}^{\infty}(1+q^{m-1/2}),\\
&&\vartheta^0_1(it)=q^{-1/48}\prod_{m=1}^{\infty}(1-q^{m-1/2}),\\
&&\vartheta^1_0(it)=\sqrt{2}q^{1/24}\prod_{m=1}^{\infty}(1+q^m),
\eea
with $q=e^{-2\pi t}$.
Here the first term and the second term in $Z_{\rm NS}$ comes from
$\Tr q^{-H}$ and $\Tr q^{-H}(-)^F$ respectively. The first term in
$Z_{\rm R}$ comes from $\Tr q^{-H}$. The final term in both
$Z_{\rm NS}$ and $Z_{\rm R}$ are due to $\Tr q^{-H}\Omega(1+(-)^F)$.
Note also that the final term in each part contributes as the NS-NS
source in $Z_{\rm NS}$ but as the R-R source in $Z_{\rm R}$.
Using the Jacobi's abstruse formula
\bea
\vartheta^0_0(it)^8-\vartheta^0_1(it)^8-\vartheta^1_0(it)^8=0,
\label{Jacobi}
\eea
we find $Z$ vanishes totally, as expected for a supersymmetric
theory.

The USp(32) theory is defined as reversing the charge of the
orientifold plane.
From a discussion of the action of $\Omega$ on open string states, we
find that the gauge group is restricted only to the SO type and the
USp type and they are related by reversing the orientifold charge
\cite{string}.
Note that this discussion holds generally and we have to reverse both
the NS-NS charge and the R-R charge of the orientifold charge.
Correspondingly, we have to add 32 anti-D9-branes for cancellation of
the R-R charge.

Thus, if we would like to read off the spectrum of the USp(32) string
theory from the type I theory, we have only to reverse the sign of the
last term in $Z_{\rm NS}$ (\ref{ZNS}), because this term corresponds
to exchange of NS-NS mode between the orientifold plane and
anti-D9-branes and only the orientifold charge is reversed.
\bea
&&Z_{\rm NS}=(32)^2\cdot\frac{\vartheta^0_0(it)^8}{\eta(it)^8}
-(32)^2\cdot\frac{\vartheta^0_1(it)^8}{\eta(it)^8}
+32\cdot\frac{\vartheta^0_1(2it)^8\vartheta^1_0(2it)^8}
{\eta(2it)^8\vartheta^0_0(2it)^8},\\
&&Z_{\rm R}=(32)^2\cdot\frac{\vartheta^1_0(it)^8}{\eta(it)^8}
-32\cdot\frac{\vartheta^0_1(2it)^8\vartheta^1_0(2it)^8}
{\eta(2it)^8\vartheta^0_0(2it)^8}.
\eea

Now we have acquired enough information to analyze the trace formula.
All we have to do is to calculate $\left.(\p/\p q)Z\right|_{q=1}$.
Using the Jacobi's abstruse formula (\ref{Jacobi}), we find
\bea
Z=2\cdot 32\cdot\frac{\vartheta^0_1(2it)^8\vartheta^1_0(2it)^8}
{\eta(2it)^8\vartheta^0_0(2it)^8}.
\eea
Since the modular forms are made of infinite polynomials, to evaluate
the trace formula, we have to use the modular transformation of the
eta function and the theta functions:
\bea
&&\eta(it)=\frac{1}{\sqrt{t}}\eta(i/t),\\
&&\vartheta^\alpha_\beta(it)=\vartheta^\beta_\alpha(i/t),
\eea
to transform the partition function $Z$ into the sum of finite
polynomials and infinite non-perturbative effects like
$e^{-2\pi/t}$.
Using the modular transformation we find
\bea
Z=2\cdot 32\cdot\sqrt{2t}^8
\frac{\vartheta^0_1(i/2t)^8\vartheta^1_0(i/2t)^8}
{\eta(i/2t)^8\vartheta^0_0(i/2t)^8}
=1024\,t^4+O(e^{-2\pi/2t}).
\eea
Especially, the coefficient of $t$ in the partition function $Z$ is
zero.
This shows that the trace formula holds and implies that in the
USp(32) theory the supersymmetry is broken spontaneously.
Note that the trace formula also holds for any power except 4.
This is the same situation as the D-brane anti-D-brane system
\cite{Yon}.

\section{Conclusion and discussion}
In this work we showed that the trace formula also holds in the
USp(32) string theory.
This suggests that the supersymmetry is broken spontaneously.
Since the supersymmetry of this system is broken by the orientifold
plane and anti-D9-branes, we expect that the orientifold plane and
anti-D9-branes are intrinsic objects of the theory.

Although this interpretation seems plausible, we cannot regard our
results as a rigorous evidence for it.
Here we shall discuss the possibility of interpreting the orientifold
plane as an intrinsic object by counting the Goldstinos which are
expected when the supersymmetry is broken spontaneously.\footnote{We
are grateful to S.\ Sugimoto for a valuable discussion on this point.}

Let us first consider the type I theory.
The type I theory is defined by adding to the type II theory an
orientifold plane and D9-branes to break half the supersymmetries.
If the orientifold plane and D9-branes are both intrinsic objects, one
should regard the supersymmetry as broken spontaneously and expect the
Goldstino to appear.
However, we cannot find a massless fermion singlet to be regarded as
the Goldstino.

In the case of the USp(32) string theory the supersymmetries are
completely broken by both the orientifold plane and anti-D9-branes.
Since we all know that anti-D9-branes break only half the
supersymmetries, the rest of the supersymmetries should be broken by
the orientifold plane.
There is a massless fermion singlet to be identified as the Goldstino.
This, however, is not enough, since originally we are considering the
type II theory.

Note also that the absence of the Goldstinos does not mean directly
that the orientifold plane is not an intrinsic object.
In fact when the broken symmetry is a local symmetry, the Goldstino
may be eaten up by some fields.
For example, when the superstring theory is compactified on the
Calabi-Yau manifold to construct a realistic model, we does not expect
to find a Goldstino.

Therefore, one might possibly stick to considering that the
orientifold plane breaks half the supersymmetries explicitly and
regard our proof of the trace formula in this paper only as a sign of
the fact that D-branes and anti-D-branes are intrinsic objects.
To avoid this possibility it remains to clarify the mechanism how the
Goldstino for the orientifold plane is eaten up by other fields.
This is an interesting future problem.

\bigskip

{\bf Acknowledgment}\\
We would like to thank H.\ Hata, S.\ Kawamoto, T.\ Matsuo and
especially S.\ Sugimoto for valuable discussions and comments.
The author is also grateful to his colleagues for their encouragement.
This work is supported in part by Grant-in-Aid for Scientific Research
from Ministry of Education, Science, Sports and Culture of Japan
(\#04633). The author is supported in part by the Japan Society for
the Promotion of Science under the Predoctoral Research Program.

\newcommand{\J}[4]{{\sl #1} {\bf #2} (#3) #4}
\newcommand{\AP}{Ann.\ Phys.\ (N.Y.)}
\newcommand{\MPL}{Mod.\ Phys.\ Lett.}
\newcommand{\NP}{Nucl.\ Phys.}
\newcommand{\PL}{Phys.\ Lett.}
\newcommand{\PR}{Phys.\ Rev.}
\newcommand{\PRL}{Phys.\ Rev.\ Lett.}
\newcommand{\PTP}{Prog.\ Theor.\ Phys.}
\newcommand{\hep}[1]{{\tt hep-th/{#1}}}


\begin{thebibliography}{99}
\bibitem{BanSus}
T.\ Banks and L.\ Susskind,
``Brane-Antibrane Forces'',
\hep{9511194}.
\bibitem{Sen}
A.\ Sen, ``Supersymmetric World-Volume Action for Non-BPS D-branes'',
\J{JHEP}{9910}{1999}{008}, \hep{9909062}.
\bibitem{Yon}
T.\ Yoneya,
``Spontaneously Broken Space-Time Supersymmetry in Open String Theory
without GSO Projection'',
\J{\NP}{B576}{2000}{219}, \hep{9912255}.
\bibitem{Sug}
S.\ Sugimoto,
``Anomaly Cancellations in the Type I D9-anti-D9 System and the
USp(32) String Theory'',
\J{\PTP}{102}{1999}{685}, \hep{9905159}.
\bibitem{SW}
J.H.\ Schwarz and E.\ Witten,
``Anomaly Analysis of Brane-Antibrane Systems'',
\J{JHEP}{0103}{2001}{032}, \hep{0103099}.
\bibitem{FGP}
S.\ Ferrara, L.\ Girardello and F.\ Palumbo,
``A General Mass Formula in Broken Supersymmetry'',
\J{\PR}{D20}{1979}{403}.
\bibitem{WB}
J.\ Wess and J.\ Bagger,
``Supersymmetry and Supergravity'', Chapter 8,
Princeton University Press.
\bibitem{DMM}
K.R.\ Dienes, M.\ Moshe and R.C.\ Myers,
``String Theory, Misaligned Supersymmetry, and the Supertrace
Constraints'',
\J{\PRL}{74}{1995}{4767}, \hep{9503055}.
\bibitem{DudMou}
E.\ Dudas and J.\ Mourad,
``Consistent Gravitino Couplings in Non-Supersymmetric Strings'',
\hep{0012071}.
\bibitem{CM}
G.F.\ Chapline and N.S.\ Manton,
``Unification of Yang-Mills Theory and Supergravity in
Ten-Dimensions'',
\J{\PL}{B120}{1983}{105}.
\bibitem{string}
J.\ Polchinski,
``String Theory'', Chapter 6, 7 and 10,
Cambridge Press.
\end{thebibliography}
\end{document}